# Diffuse phase transition and electrical properties of lead-free piezoelectric $(Li_xNa_{1-x})NbO_3$ ($0.04 \leq x \leq 0.20$) ceramics near morphotropic phase boundary

S. Mitra, A. R. Kulkarni, Om. Prakash

*Department of Metallurgical Engineering and Materials Science, Indian Institute of Technology Bombay, Mumbai 400076, India*

**ABSTRACT:**

Temperature-dependent dielectric permittivity of lead-free $(Li_xNa_{1-x})NbO_3$ for nominal $x = 0.04$-$0.20$, prepared by solid state reaction followed by sintering, was studied to resolve often debated issue pertaining to exactness of morphotropic phase boundary (MPB) location along with structural aspects and phase stability in the system near MPB. Interestingly, a diffuse phase transition has been observed in the dielectric permittivity peak arising from the disorder induced in *A*-site and structural frustration in the perovskite cell due to Li substitution. A partial phase diagram has been proposed based on temperature-dependent dielectric permittivity studies. The room temperature piezoelectric and ferroelectric properties were investigated and the ceramics with $x = 0.12$ showed relatively good electrical properties ($d_{33} = 28$ pC/N, $k_p = 13.8\%$, $Q_m = 440$, $P_r = 12.5$ μC/cm$^2$, $E_C = 43.2$ kV/cm, $T_m = 340$ °C). These parameter values make this material suitable for piezoelectric resonator and filter applications. Moreover, a high dielectric permittivity ($\varepsilon'_r = 2703$) with broad diffuse peak near transition temperature along with low dielectric loss (< 4%) in a wide temperature range (50-250 °C) found in this material may also have a potential application in high-temperature multilayer capacitors in automotive and aerospace related industries.

## I. INTRODUCTION

Research on lead free piezoelectric materials has advanced with increasing efforts in recent times to find suitable replacements for the existing Pb(Zr,Ti)O$_3$ (PZT) materials, due to the environmental concern of toxic lead (Pb). Reports on the structure-property correlations of many Pb-free piezoelectric ceramic systems have since came along. Amongst the Pb-free materials, alkali niobates have been found attractive due to good piezoelectric properties and high Curie temperature. Promising piezoelectric properties in alkali niobates are believed to arise from the coexistence of thermodynamically equivalent phases at the morphotropic phase boundary (MPB) composition. At MPB, enhanced piezoelectric and electromechanical responses are found due to the ease of polarization rotation under the external electric field [1]. The concept of MPB (separating orthorhombic (O) and tetragonal (T) phases at $x = 0.12$) in the lead-free (Li$_x$Na$_{1-x}$)NbO$_3$ (commonly known as LNN) system was first introduced by Zeyfang *et. al*, in analogy with PZT, on the basis of the highest obtained piezoelectric properties [2,3]. Unlike PZT [4], NaNbO$_3$, as an end member of LNN system, has orthorhombic symmetry (*Pbma*) and is antiferroelectric while the other end member LiNbO$_3$ has a rhombohedral symmetry (*R3c*) and is ferroelectric. MPB in LNN system is reported to exist at $x = 0.12$. However, on the subject of MPB, different perspectives still exist on the exactness of $x$ value and character of the phases in proximity. Zeyfang *et. al* introduced a partial phase diagram [3,5], following Nitta *et. al* [6], showing a phase transformation from O to T phase at $x = 0.12$. In contrast, many authors have reported the coexistence of orthorhombic (O) and rhombohedral (R) phase near $x = 0.12$ [7-13]. While, Norbe *et. al* reported continuity of a single O phase composition up to $x = 0.12$ prepared using Polymeric precursor method [14]. As a result of this disagreement, it is suspected that the phases in LNN ceramics near MPB are very much process sensitive. Therefore, possibility behind this discrepancy could be due to deviation of sample composition from the aimed stoichiometry in the final product, for the alkali elements Li, Na which have high vapour pressure, and some volatility at sintering temperatures. Many techniques are available to minimize the effect of volatilization for example, spark plasma sintering [15], double crucibles [16], which are perplexed and not desirable for mass production in industries. In addition to this, the use of excess alkali elements to compensate for the loss was also conceived as an alternative, however, excess alkali addition promotes grain growth which deteriorates dielectric properties in the resulting ceramic [17]. In view of the above, our strategy is to carry out quantitative compositional analysis of the sintered product as a necessary step to determine the exact composition. This would also help in correctly interpreting the measured electrical properties near the MPB compositions.

In this paper, we have investigated the structural aspects of LNN ($x$ = 0.04-0.20) ceramics, prepared by conventional ceramics route, using XRD data and Raman spectroscopy and an MPB has been located in the nominal LNN composition near $x$ = 0.14, where orthorhombic symmetry gradually changes to rhombohedral symmetry at room temperature. A drop in $x$ away from the actual MPB composition is anticipated due to some volatility especially that of Li and to some extent of Na. Here, a partial phase diagram has been proposed based on temperature dependent dielectric permittivity studies. Further, a very interesting feature, a Diffuse Phase Transition (DPT), which may furnish a new insight into developing high-temperature lead-free ferroelectrics with DPT as electrostrictive materials [18], is observed in dielectric permittivity peak, and degree of diffuseness is found more pervasive in the MPB compositions. Therefore, it is significant to understand the origin of DPT in LNN ceramics industrially as well as scientifically. In here, some of the ferroelectric and piezoelectric properties of the ceramic samples near MPB composition are also reported.

## II. EXPERIMENTAL

All the LNN samples in this study were prepared using conventional solid state reaction followed by sintering. The starting raw materials were reagent-grade $Nb_2O_5$, $Na_2CO_3$ (both 99.5% pure, Loba Chemie, India) and $Li_2CO_3$ (99.0% pure, Merck, India). These were mixed in the desired stoichiometry of $Na_{1-x}Li_xNbO_3$ ($x$ = 0.04-0.20), and wet ball milled in alcohol to obtain proper mixing and a surface-active fine powder. The finely mixed constituent powders were then solid state reacted (900 °C/5h), followed by compaction into pellet and pressureless sintering at different temperatures 1180-1260 °C for 2h. The phase purity of the ceramic samples was checked by powder X-Ray Diffraction (XRD) at room temperature using X-ray diffractometer (X'Pert, PANalytical) with Cu-Kα radiation. For this, the final sintered pellets were finely crushed and annealed at 500 °C for 12h to reduce residual strain introduced by crushing. The details of chemical homogeneity and a possible deviation from desired stoichiometry in the final product due to the use of volatile alkali elements Li, Na were checked by Inductively Coupled Plasma-Atomic Emission Spectroscopy (ICP-AES, SPECTRO ARCOS, Germany). For this, sintered samples were dissolved using microwave-assisted acid (mixture of aqua regia and HF) dissolution techniques [19] in a Microwave-digestion system (MILESTONE START D) at 200 °C for 3h. Raman spectra were collected in the range of 100-1000 cm$^{-1}$ at room temperature using 514.5 nm excitation from an Ar$^+$ laser using Jobin-Yvon LabRAM HR800 (Horiba). The dielectric constant/permittivity of the unpoled samples was recorded using Impedance Analyzer (Alpha High Resolution, Novocontrol, Germany) in the frequency range 1-100 kHz over the temperatures 50-500 °C. The *P-E* hysteresis loop was traced using ferroelectric analyzer (TF Analyzer 2000, Aixact, Germany) at 1 Hz. The planar electromechanical coupling factor $k_p$ and mechanical quality factor $Q_m$ were determined on the poled samples using impedance analyzer via resonance-antiresonance method [20]. For this, the disk samples were poled at 150°C under the DC field of 5-8 kV/mm. The piezoelectric constant $d_{33}$ was measured using $d_{33}$-meter (Piezotest, PM300, UK).

## III. RESULTS AND DISCUSSION

### A. Phase Analysis

Figure 1 shows the XRD pattern of LNN ceramics for different values of composition $x$ (i.e., Li content) at room temperature. All compositions show a characteristics pure perovskite structure free from the presence of any secondary phases. An amplified view near $2\theta \sim 47°$ shows presence of a major O phase (JCPDS # 033-1270) for $x \leq 0.12$, characterized by splitting of (202)/(080) peaks. As $x$ increases (for $x \geq 0.14$), (202)/(080) peaks merged into a single peak (024), along with the appearance of superstructure peaks characteristics of $R3c$ symmetry, indicating evolution of a R phase (JCPDS # 53-0341). An MPB therefore may exist approximately at $x = 0.14$, indicating the coexistence of both O and R phases in agreement to other reports [9-11].

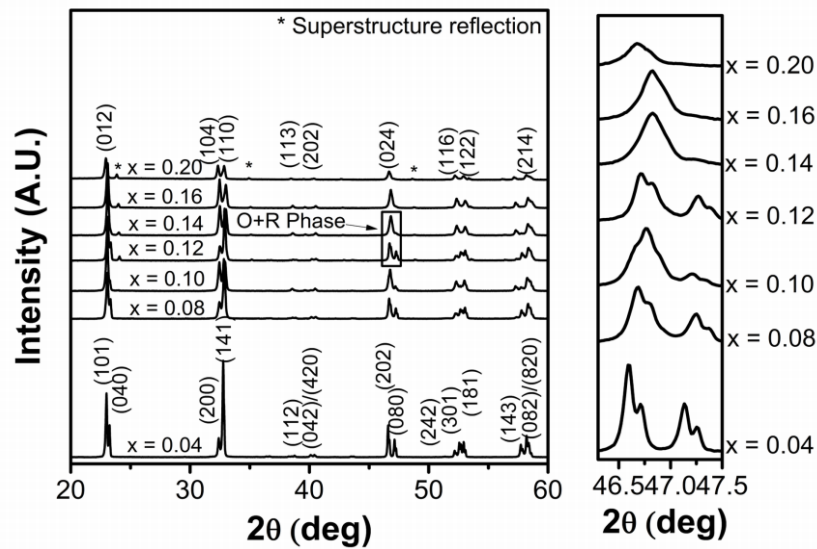

Figure 1: X-ray diffraction pattern of $Na_{1-x}Li_xNbO_3$ ($x = 0.04$-$0.20$) ceramics; amplified XRD pattern in the $2\theta$ range 45.5°-47.5° shown in the right side.

Different phases with different distortions extend to structural modification which dramatically changes the physical properties. Therefore, to look into the cell distortions in perovskite LNN ceramics, it is expedient to consider the normalized cell parameter, $a' = V^{1/3}$, where $V$ is the volume per unit formula unit of the reduced cubic perovskite cell [21]. Figure 2(a) shows the variation of normalized cell parameter with $x$ (Li content). The observed scattered values of cell parameter ($a'$) for $x \leq 0.12$ indicate that the smaller Li$^+$ may occupy the tetrahedral void along with the vacant lattice site, forming interstitial solid solution. Thereafter, for $x \geq 0.12$, the cell parameter changes almost linearly with $x$, confirming a substitution solid solution following Vegard's law [22, 11].

The stability of perovskite structure can be defined in terms of the tolerance factor ($t$) values proposed by Goldschmidt *et al.* [23], and calculated using Shannon's radii [24] for LNN ceramics is shown in Figure 2(b). The tolerance factor of LNN ceramics found to be 0.968 for $x = 0.04$ and decreases to 0.949 for $x = 0.20$, indicating a change from higher symmetry (orthorhombic, *Pbma*) to a lower symmetry (rhombohedral, *R3c*). The *t*-values ranging from 0.968 to 0.949 ($0.8 < t < 1.0$) confirms that structure is stable and does not show tetragonal symmetry at room temperature for the composition range studied [25, 26]. However, some extent of instability in the structure arises due to size mismatch of Li$^+$ ($r_{Li+}$ = 1.26 Å) and Na$^+$ ($r_{Na+}$ =1.39 Å) for 12 coordination [24] which induces a disorder in *A*-site cation. The *A*-site cation has a strong interaction with oxygen anion in ABO$_3$-structure and a disorder in *A*-site perturbs the crystal symmetry. This leads to a structural transformation from orthorhombic to rhombohedral symmetry due to large distortion caused by Li$^+$ as confirmed by Raman spectra.

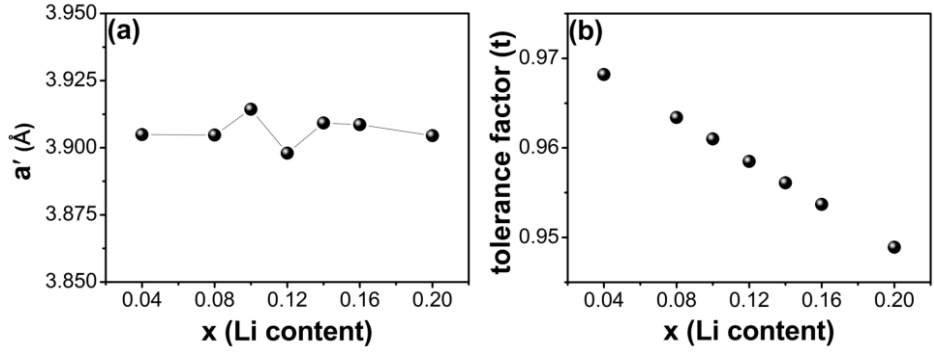

Figure 2: (a) Variation of normalized cell parameter $a′$ of the reduced cubic perovskite cell, (b) Variation of tolerance factor ($t$) for $Na_{1-x}Li_xNbO_3$ ceramics as a function of Li content.

Raman spectroscopy has been used as a successful tool to investigate structure of $NaNbO_3$, $LiNbO_3$ and LNN [9]. The Raman spectra of LNN ($x$ = 0.04-0.20) ceramics are shown in Figure 3. The vibrational modes of LNN ceramics can be categorized as internal modes of $NbO_6$ octahedron and the lattice translational mode involving $Na^+$ and or $Li^+$ cation motion against $NbO_6$ octahedron. The vibrational modes of an isolated $NbO_6$ octahedron can be separated into two pure bond stretching vibrations such as: Nb-O stretching vibrations of $A_{1g}(\nu_1)$ and O-Nb-O stretching vibrations of $E_g(\nu_2)$ symmetry, two interbond angle bending vibrations of $F_{2g}(\nu_5)$ and $F_{2u}(\nu_6)$ symmetry whereas, $\nu_3$ and $\nu_4$ vibrations are considered as a combination of stretching and bending of $F_{1u}$ symmetry [27]. The rotational modes of $NbO_6$ octahedron and translational mode of $Li^+/Na^+$ are assigned between 100-200 $cm^{-1}$ and the other bands (200-1000 $cm^{-1}$) are accorded with internal vibrations of $NbO_6$ octahedron.

The high frequency vibrational modes found near 615 $cm^{-1}$ and 570 $cm^{-1}$ in the LNN Raman spectra are attributed as $\nu_1$ and $\nu_2$ stretching mode in $NbO_6$, respectively whose position almost remain same either in $LiNbO_3$ (rhombohedral, *R3c*) or $NaNbO_3$ (orthorhombic, *Pbma*) side (Figure 3(b)). However, a broadening of $\nu_1$ peak for $x \geq 0.14$, suggests a structural disorder due to substitution of Li at *A*-site. The broadening of peak near 251 cm$^{-1}$ assigned as $\nu_5$ bending mode of NbO$_6$, increases as the Li concentration is increasing (from $x = 0.04$ to $0.20$), also indicating introduction of a distortion in NbO$_6$ octahedra (Figure 3(a)). The peak near 188 cm$^{-1}$ assigned as $\nu_6$ shifts to the higher frequency and disappears, while a peak near 230 cm$^{-1}$ starts evolving for $x \geq 0.14$ (Figure 3(b)), suggests a gradual changing of O to R phase. The intensity of the peak near 117 cm$^{-1}$ decrease as Li concentration is increased (Figure 3(a)) and also shifts to the lower frequency for $x \geq 0.14$ (Figure 3(b)) suggesting *A*-site cation displacement against the oxygen-octahedra.

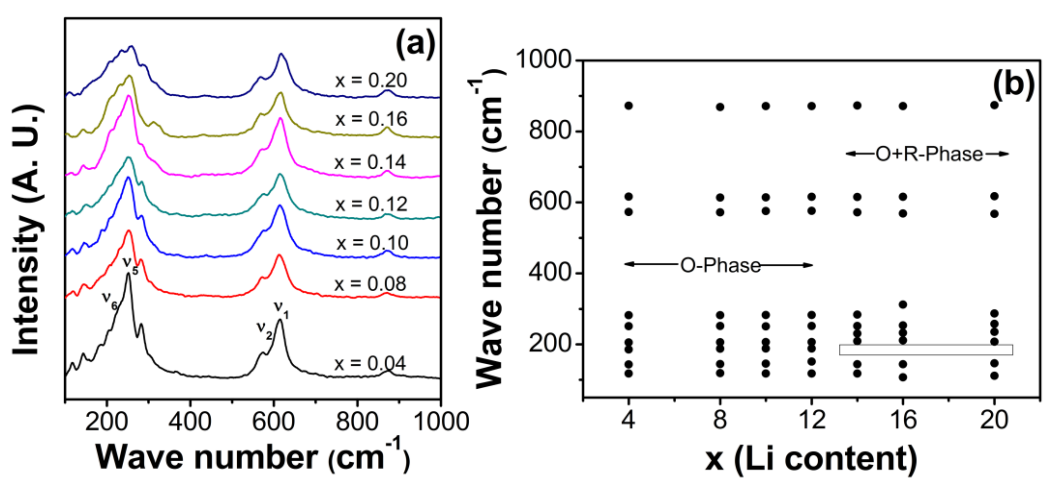

Figure 3: (Colour online) (a) Raman spectra, (b) Frequencies of the Raman peaks as a function of Li content (*x*) in for Na$_{1-x}$Li$_x$NbO$_3$ ceramics at room temperature.

## B. Compositional Analysis

In order to check the possibility of deviation from the desired stoichiometry due to deficiency of alkali elements in the sintered bulk LNN ceramics, ICP-AES analysis has been carried out for all compositions ($x = 0.04$-$0.20$). The comparison of concentration of each element in LNN between the nominal and ICP-AES measured final sintered ceramics (in mol%) with an accuracy of 0.05 mol% is made in Table 1. A significant deviation from the desired stoichiometry has been encountered in nominal and the final sintered product which arises due to the volatile nature of alkali elements at elevated temperature. A drop in *x*, about 2 mol% away from the nominal composition for $x$ = 0.04-0.14 while ~3 mol% for $x$ = 0.16 and 0.20, has been observed in the final sintered product.

**C. Dielectric permittivity**

The real and imaginary part of permittivity, ($\varepsilon'_r$) and ($\varepsilon''_r$) vs temperature ($T$) plot at different frequencies (1-100 kHz) for LNN ceramics are shown in Figure 4. Both $\varepsilon'_r$ and $\varepsilon''_r$ show two peaks which represent a ferroelectric to paraelectric transition at $T_m$ and ferroelectric to ferroelectric transition at $T_{ff}$, appears as a shoulder at lower temperature side of permittivity maxima. A broadening in the permittivity peak near both transitions has been observed, however, there was no frequency dispersion in $T_m$. Moreover, the temperature corresponds to the maximum in $\varepsilon'_r(T_m)$ found higher to that of $\varepsilon''_r(T_m)$ for all compositions. All such behaviour is associated with the diffuse phase transition (DPT) in ferroelectric systems and differs from relaxor ferroelectrics. The degree of diffuseness is normally measured from the diffuseness parameter $\gamma$, using modified Curie-Weiss law: $1/\varepsilon'_r - 1/\varepsilon'_m = C^{-1}(T-T_m)^\gamma$, where $\varepsilon'_m$ is the value of maximum permittivity at the phase transition temperature $T_m$. The diffuseness parameter $\gamma$ has a value 1 for ideal ferroelectrics and 2 for relaxor [28].

From the slope of $\ln(1/\varepsilon'_r - 1/\varepsilon'_m)$ vs $\ln(T-T_m)$, diffuseness parameter $\gamma$ can be determined from the $\varepsilon'_r(T)$ data at 1 kHz as can be seen in Figure 5. The inset of Figure 5 shows variation of $\gamma$ with $x$ (Li content) and found to increase from ~1.2 for $x$ = 0.04 to ~1.8 for $x$ = 0.14 and then starts decreasing. The high values of $\gamma$ near MPB composition (for $x$ ~ 0.14), are believed to arise from a structural frustration and high degree of disorder in $A$-site, induced by octahedral distortion in $R3c$-rich phase due to partial Li substitution, and responsible for this DPT behaviour. The cationic disorder at $A$-site gives rise to a random local field and develops a long range polar ordering, resulting in polar nano regions (PNRs) with local $T_C$ each [29, 30]. As the temperature increases thermal fluctuations of the local field increase giving rise to large distribution of local Curie temperature and as a consequence broadening in the dielectric peak observed [31].

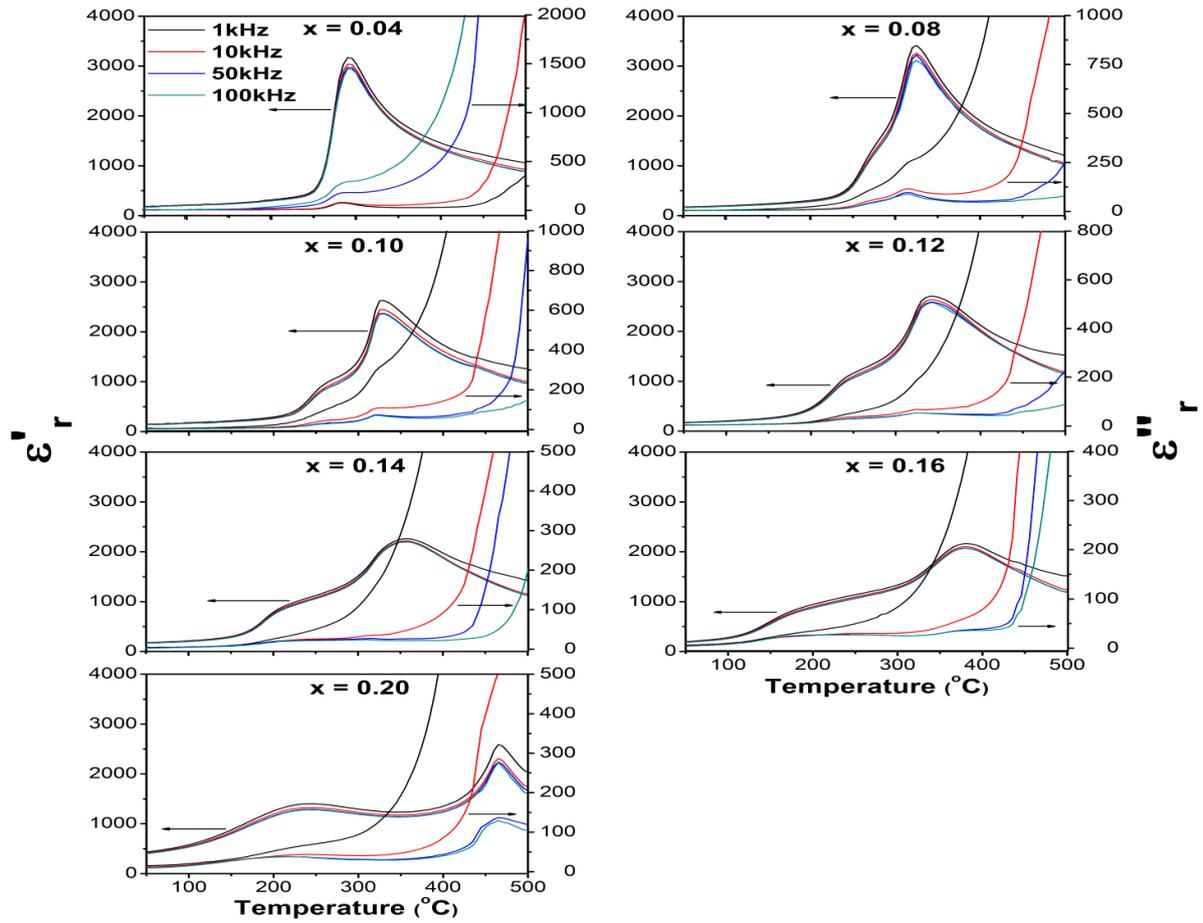

Figure 4: (Colour online) Temperature dependence of real and imaginary part of dielectric permittivity at various frequencies for $Na_{1-x}Li_xNbO_3$ ($x$ = 0.04-0.20) ceramics.

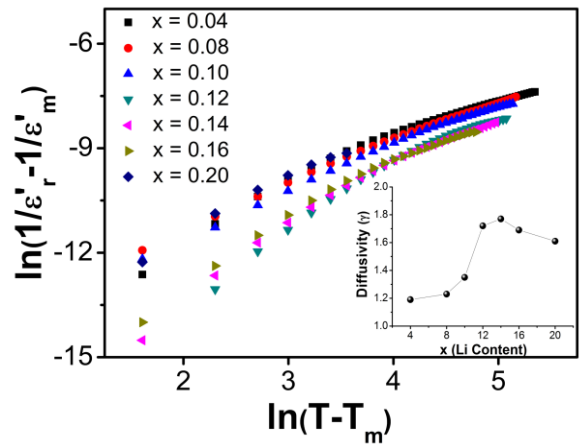

Figure 5: (Colour online) Plot of ln($1/\varepsilon'_r - 1/\varepsilon'_m$) vs ln($T-T_m$); inset showing the variation of degree of diffuseness ($\gamma$) as a function of Li content.

The variations of $T_m$ and $T_{ff}$ with $x$ are summarized to form a partial phase diagram, as shown in Figure 6, where the transition temperatures ($T_m$ and $T_{ff}$) are determined from the maximum of the first derivative of permittivity at which permittivity changes strongly [32]. For $x = 0.04$, two peaks are not observed as the two phase transitions are very close to each other. As $x$ increases from 0.08 to 0.14, transition peak $T_m$ found to shift towards higher temperature and $T_{ff}$ gradually shifts towards lower temperature. Further increase in $x$, causes $T_{ff}$ and $T_m$ both to shift towards higher temperature side.

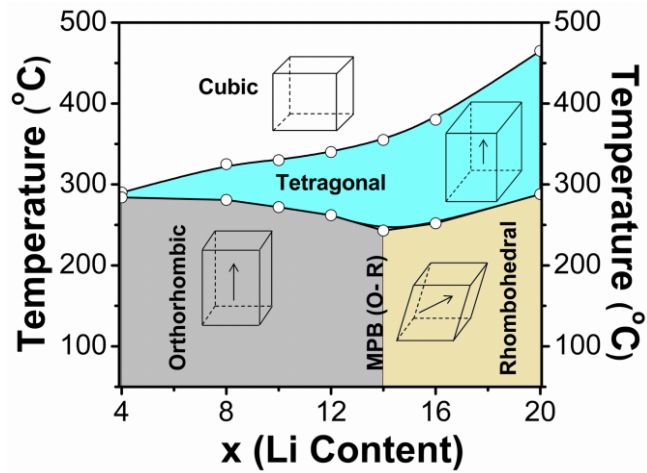

Figure 6: (Colour online) Temperature-composition phase diagram of $Na_{1-x}Li_xNbO_3$ ($x = 0.04$-$0.20$) ceramics, based on dielectric permittivity. An MPB found at (near) nominal $x = 0.14$, separating orthorhombic and rhombohedral phases.

The results, together with XRD and Raman spectroscopic analysis, support that the symmetry of the ceramics is orthorhombic for $x \leq 0.12$ and progressively changed to rhombohedral for $x \geq 0.14$ at room temperature. The location of MPB, which is found to exist around $x = 0.14$ with a vertical temperature line suggests a wide range of temperature stability [33, 34]. However, to understand mechanisms of these phase transitions, further investigation such as high-temperature structural analysis is necessary.

**D. Ferroelectric hysteresis**

The room temperature P-E hysteresis loop measured on unpoled samples under an applied electric field of 60 kV/cm at 1 Hz is shown in Figure 7(a) for comparison. All the samples show a normal hysteresis loop confirming ferroelectric nature of LNN for all compositions. The inset in Figure 7(a) shows remnant polarization $P_r$ and coercive field $E_c$ of the samples as a function of $x$. The $P_r$-values of the ceramics for $x = 0.04$ is found to be 1.8 $\mu C/cm^2$ and increases upto 10.9 $\mu C/cm^2$ for $x = 0.16$. Further increase in $x$ causes a decrease in $P_r$-value. This could be related to the nature of phase transition in the system. The strong polarization in the rhombohedral-rich regions ($x \geq 0.14$) can be realized due to Na displacement by large off centering nature of Li in the *R3c*-rich region and therefore found to have a maximum value (10.9 $\mu C/cm^2$ for $x = 0.16$) where distortion in the cell is maximum. The $E_c$-value is also found to increases from 25.1 to 34.5 kV/cm as $x$ increases from 0.04 to 0.16 and then decreases with further increase in $x$. It is due to the presence of internal bias field, which can be asserted from the asymmetric shape and horizontal off-set of coercive field, which essentially increase the $E_c$-value. $E_c$ may also increases due to large distortions in the *R3c*-rich compositions by decreasing volume of switchable domains. For the ceramics with $x = 0.12$-0.16, it was found to withstand a very high applied field (90 kV/cm) as can be seen in the *P-E* hysteresis loop shown in Figure7(b). The corresponding highest remnant polarization, $P_r = 17.2$ $\mu C/cm^2$ was observed under the applied electric field 90 kV/cm at 1 Hz for $x = 0.16$.

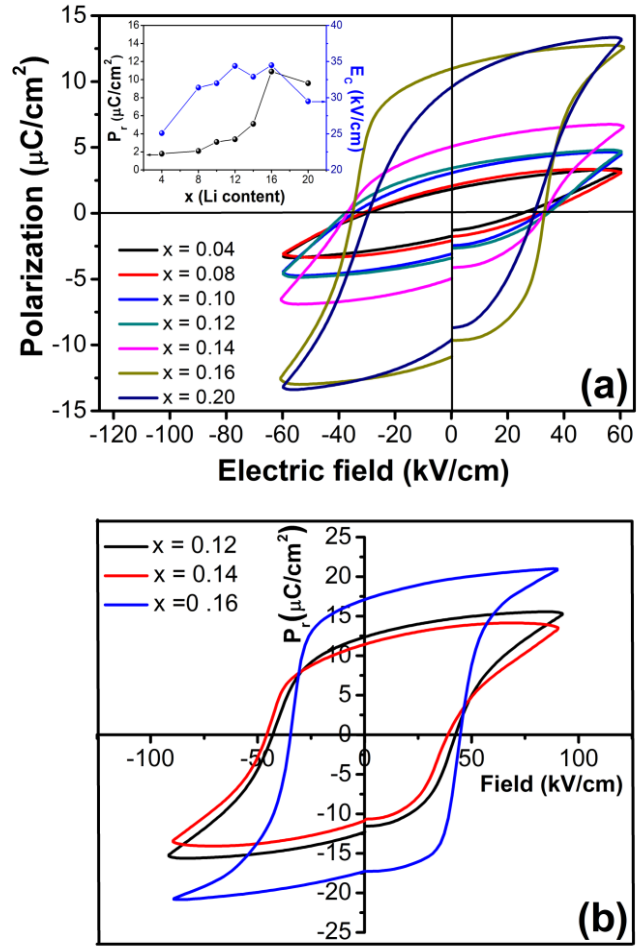

Figure 7: (Colour online) (a) P-E hysteresis loops of $Na_{1-x}Li_xNbO_3$ ($x$ = 0.04-0.20) ceramics. The inset shows $P_r$ and $E_c$ as a function of $x$; (b) P-E hysteresis loops of $Na_{1-x}Li_xNbO_3$ ($x$ = 0.12-0.16) ceramics under applied field of 90 kV/cm.

**E. Piezoelectric properties**

Figure 8 shows the piezoelectric constant ($d_{33}$), planar coupling constant ($k_p$), and mechanical quality factor ($Q_m$) of LNN ceramics ($x$ = 0.04-0.20) at room temperature. The maximum $Q_m$ value of 497 was found for $x$ = 0.08 and decreases with further increase in Li-content ($x$). The higher $Q_m$-values can be related to the low spontaneous polarization in the unit volume and hence low value $P_r$ -values as can be seen the inset of Figure 7(a). Consequently, the internal friction induced by the rotation of spontaneous polarization vector or mobility of domain wall decreased. The maximum $d_{33}$ values of 38 pN/C was found for $x$ =

0.10, whereas a maximum $k_p$ value of 21.6% for x = 0.16. Both $d_{33}$ and $k_p$ values are found not changes much with other values of Li-content.

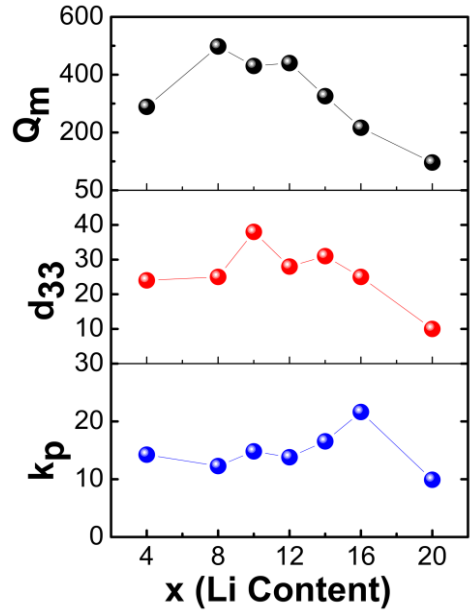

Figure 8: (Colour online) $Q_m$, $d_{33}$ and $k_p$ values as a function of Li content for $Na_{1-x}Li_xNbO_3$ ceramics.

The dielectric, ferroelectric and piezoelectric properties of LNN ceramics for $x$ = 0.04-0.20 are summarized in Table 2. It can be concluded that a large distortion in the lattice near MPB are responsible for increase in $P_r$ and $E_c$ values, however, a deviation from stoichiometry resulting from the volatile nature of Li and Na may creates defects which restrict the domain wall mobility and results in reduction in maximum dielectric constant ($\varepsilon'_r$), dielectric loss ($tan\delta$) and piezoelectric activity while coercive field ($E_C$) and mechanical quality factor ($Q_m$) have increased.

**IV. CONCLUSIONS**

In conclusion, additional structural evidences have been gathered from X-ray diffraction combined with Raman spectroscopy. Compositional analysis using ICP spectroscopy helped in locating the exact MPB between O and R phases at $x \sim 0.14$, which agrees with other reports, bringing out the clarity on MPB. A DPT observed for the compositions near MPB arises from a structural frustration induced due to an equivalent site in the lattice which is occupied by different types of cations. A partial phase diagram has been proposed based on temperature dependent dielectric permittivity data. Ferroelectric and piezoelectric properties are investigated for compositions near MPB. For the ceramics with $x = 0.12$ exhibit relatively good electrical properties ($d_{33} = 28$ pC/N, $k_p = 13.8\%$, $Q_m = 440$, $P_r = 12.5$ μC/cm$^2$, $E_C = 43.2$ kV/cm, $T_m = 340$ °C). Therefore, high temperature stability, high Curie temperature along with high mechanical quality factor makes this material suitable for piezoelectric resonator and filter applications. In addition, ceramics with $x = 0.12$ also shows a broad permittivity peak with maximum value $\varepsilon'_r = 2703$ and low dielectric loss ($< 4\%$) in a broad temperature range (50-250 °C) which may have a potential application in high-temperature multilayer capacitors in automotive industries.

## ACKNOWLEDGEMENT:

The authors would like to thank SAIF, IIT Bombay for carrying out ICP-AES and room temperature Raman spectroscopy and Prasanta Kumar Ojha for $d_{33}$ measurements.

TABLE I. Comparison of elements in $Na_{1-x}Li_xNbO_3$ ($x$ = 0.04-0.20) ceramics between nominal and ICP-AES measured sintered product values in mol%.

| Li content ($x$) | Nominal (mol%) | | | Measured in sintered product (mol%) | | |
|---|---|---|---|---|---|---|
| | Li | Na | Nb | Li | Na | Nb |
| 0.04 | 4 | 96 | 100 | 2.39 | 95.77 | 100.00 |
| 0.08 | 8 | 92 | 100 | 6.29 | 91.53 | 100.27 |
| 0.10 | 10 | 90 | 100 | 8.53 | 90.65 | 99.82 |
| 0.12 | 12 | 88 | 100 | 10.07 | 89.25 | 100.13 |
| 0.14 | 14 | 86 | 100 | 12.07 | 86.75 | 100.21 |
| 0.16 | 16 | 84 | 100 | 13.18 | 83.94 | 99.88 |
| 0.20 | 20 | 80 | 100 | 17.28 | 79.85 | 100.29 |

TABLE II. Dielectric, ferroelectric and piezoelectric properties of of $Na_{1-x}Li_xNbO_3$ ($x$ = 0.04-0.20) ceramics

| Li content ($x$) | $\varepsilon'_r$ ($T_m$) | $tan\delta$ (RT) | $P_r^\dagger$ ($\mu C/cm^2$) | $E_C^\dagger$ (kV/cm) | $Q_m$ |
|---|---|---|---|---|---|
| 0.04 | 3167 | 0.042 | - | - | 288 |
| 0.08 | 3412 | 0.029 | - | - | 497 |
| 0.10 | 2624 | 0.026 | - | - | 430 |
| 0.12 | 2703 | 0.026 | 12.5 | 43.2 | 440 |
| 0.14 | 2261 | 0.027 | 11.3 | 38.9 | 326 |
| 0.16 | 2166 | 0.030 | 17.2 | 45.0 | 216 |
| 0.20 | 2589 | 0.039 | - | - | 96 |

† Measured at 90 kV/cm, $T_m$: Temperature of highest permittivity, RT: Room temperature

**List of figure captions**

**Figure 1:** X-ray diffraction pattern of $Na_{1-x}Li_xNbO_3$ ($x$ = 0.04-0.20) ceramics; amplified XRD pattern in the 2θ range 45.5°-47.5° shown in the right side.

**Figure 2:** (a) Variation of normalized cell parameter $a′$ of the reduced cubic perovskite cell, (b) Variation of tolerance factor ($t$) for $Na_{1-x}Li_xNbO_3$ ceramics as a function of Li content.

**Figure 3:** (Colour online) (a) Raman spectra, (b) Frequencies of the Raman peaks as a function of Li content ($x$) in for $Na_{1-x}Li_xNbO_3$ ceramics at room temperature.

**Figure 4:** (Colour online) Temperature dependence of real and imaginary part of dielectric permittivity at various frequencies for $Na_{1-x}Li_xNbO_3$ ($x$ = 0.04-0.20) ceramics.

**Figure 5:** (Colour online) Plot of $\ln(1/\varepsilon'_r - 1/\varepsilon'_m)$ vs $\ln(T-T_m)$; inset showing the variation of degree of diffuseness ($\gamma$) as a function of Li content.

**Figure 6:** (Colour online) Temperature-composition phase diagram of $Na_{1-x}Li_xNbO_3$ ($x$ = 0.04-0.20) ceramics, based on dielectric permittivity. An MPB found at (near) nominal $x$ = 0.14, separating orthorhombic and rhombohedral phases.

**Figure 7:** (Colour online) (a) P-E hysteresis loops of $Na_{1-x}Li_xNbO_3$ ($x$ = 0.04-0.20) ceramics. The inset shows $P_r$ and $E_c$ as a function of $x$; (b) P-E hysteresis loops of $Na_{1-x}Li_xNbO_3$ ($x$ = 0.12-0.16) ceramics under applied field of 90 kV/cm.

**Figure 8:** (Colour online) $Q_m$, $d_{33}$ and $k_p$ values as a function of Li content for $Na_{1-x}Li_xNbO_3$ ceramics.